# Hidden Semi-Markov Models for Single-Molecule Conformational Dynamics

A. Kovalev[*], N. Zarrabi[*], F. Werz[*], M. Börsch[*], Z. Ristic[#], H. Lill[#], D. Bald[#], C. Tietz[*], J. Wrachtrup[*]

[*]3. Physikalisches Institut, Universität Stuttgart, Pfaffenwaldring 57, 70569 Stuttgart, Germany
[#]Department of Structural Biology, Faculty of Earth and Life Science, Vrije Universiteit Amsterdam, De Boelelaan 1085, 1081 HV Amsterdam, Netherlands

**Abstract**
The conformational kinetics of enzymes can be reliably revealed when they are governed by Markovian dynamics. Hidden Markov Models (HMMs) are appropriate especially in the case of conformational states that are hardly distinguishable. However, the evolution of the conformational states of proteins mostly shows non-Markovian behavior, recognizable by non-monoexponential state dwell time histograms. The application of a Hidden Markov Model technique to a cyclic system demonstrating semi-Markovian dynamics is presented in this paper and the required extension of the model design is discussed. As standard ranking criteria of models cannot deal with these systems properly, a new approach is proposed considering the shape of the dwell time histograms. We observed the rotational kinetics of a single $F_1$-ATPase $\alpha_3\beta_3\gamma$ sub-complex over six orders of magnitude of different ATP to ADP and $P_i$ concentration ratios, and established a general model describing the kinetics for the entire range of concentrations. The HMM extension described here is applicable in general to the accurate analysis of protein dynamics.

**Introduction**

Single-molecule experiments, in comparison to ensemble measurements, allow to reconstruct the distribution of various molecular parameters. Often it is the aim to identify separate states or conformations of the molecule and to express state dynamics. Therefore the typical approach for a state definition is to introduce thresholds in order to classify data points into different co-domains and to assign these areas to different states afterwards (1,2). Once the states are identified, it is possible to derive transition rates between those states in order to characterize the dynamic behavior of the molecule. This approach works as long as the states can be separated by thresholds, but fails as soon as their co-domains overlap. In single-molecule experiments this overlap is caused by a poor signal-to-noise ratio (SNR) especially when the state duration lasts only a few milliseconds. In this case, increasing data accumulation time or higher time resolution does not improve a proper state identification.
For instance, when conformational changes of an enzyme are investigated by Förster-type fluorescence resonance energy transfer (FRET) (3,4) the average photon count rate of organic fluorophores is usually between $10^4$ and $10^5$ photons per second (5,6). With state transition rates around $10^2$ to $10^3$ s$^{-1}$, about 100 photons remain for identifying the FRET efficiency of a state within the data stream. Since the signal-to-noise ratio (SNR) depends on two fluorophore photon count rates, that is, 50 counts in each of the two channels, and the SNR reaches a value of about 5:1 .

When the fluorescent marker is replaced by a small particle observable by bright- or dark-field microscopy, its Brownian motion at room temperature blurs the particle position, which makes it difficult to follow fast conformational changes. The bead attached to the $F_1$-ATPase examined in this paper moves on average 14 nm within a 5 ms image frame during rotation. However the mean thermal displacement of the marker bead about 25 nm per frame yields a SNR of 0.6:1. SNR of single-molecule force measurements using optical tweezers (7) are of the same range. Recently, the SNR in nucleic acid translocating single-molecule motor experiments was found to be below 2.4 at transition rates above 15 $s^{-1}$ (8). Another example is the motion of single-headed kinesin KIF1A with a latex bead as a marker observed in an optical trap. Its step size is determined to 8 nm with a standard deviation of 15 nm, i.e. the SNR is 0.5:1 (9).

Additional information is required for a proper state assignment in those cases. A Hidden Markov model (HMM) utilizes the architecture of a conformational state network as *a priori* information. It assigns a separate probability density function (PDF) to each observable for every hidden state. The advantage of HMMs is that this method still assigns states correctly even if the PDFs overlap. Among many other applications, e.g. in speech recognition (10) or DNA analysis, trajectory analysis with Hidden Markov Models was successfully applied earlier to systems with inherent Markov states (11,12). As soon as the transition from the current state is dependent on the previous states of the system (i.e. non-Markovian behaviour) a standard Hidden Markov approach can not be used anymore. Nevertheless, in some special cases it is possible to incorporate system memory into the Markov approach to derive the underlying dynamic behaviour (11). Enzymes e.g. cyclically change among a finite number of conformations along catalytic reaction pathways and return to the initial condition after each reaction cycle. Therefore the memory dependence is restricted to the maximum number of conformations. Consequently, those systems can be treated as semi-Markovian by an extension of the set of conformational states (13,14).

Complex systems like proteins often have conformational states with multi-exponential dwell time distributions (15). That means, that state transitions can not be properly described by one rate constant. This is a clear feature of non-Markovian processes. An illustrative example of an enzyme with non-Markovian dynamics is the $F_1$-ATPase $\alpha_3\beta_3\gamma$ sub-complex, which is an tunable ATP-driven effective molecular motor (16-18). Like in the holoenzyme $F_oF_1$-ATP synthase (19-21) the γ subunit of this sub-complex rotates in 120° steps at millimolar ATP concentrations; sub-millisecond image analysis revealed that each $120^o$ step comprises a $80^o$ and a $40^o$ sub-step (22,23). To visualize the stepwise rotational movement of the rotary subunit a bead was bound to the γ-subunit (22). The stepwise movement of the bead is superimposed by Brownian motion, which yields a strong broadening of the expected three stopping positions. Using small beads or low rotational speeds at nanomolar ATP concentrations a separation of the three states is easily possible by applying thresholds (22). Here, using $F_1$-ATPase as a nanomotor operated under high load (i.e. with a large bead) and at high rotational speed (i.e. at high ATP concentrations) the PDFs of the bead stopping positions overlapped. Because the shape of the dwell time histograms of the $F_1$-ATPase states were found to be non-monoexponential, a state expansion approach to the Hidden Markov Model similar to (24) was used to take into account the histograms shape. A single model was developed describing the rotating behavior of the $F_1$-ATPase nanomotor at different rotational speeds.

**Material and Methods**

**Hidden Markov Models**
The framework of a HMM consists mainly of a process that generates time series $\{q_t\}$ of a finite number of not directly observable and therefore "hidden" states $q$ according to an autonomous Markov process. The transition probability to jump from state $q_i$ into another state $q_j$ is described by a time independent parameter $k_{ij}$ (for a three state model, see Figure 1 a). A model with $n$ states is therefore characterized by a state transition matrix **K** with $n \times n$ entries. This time resolution ($\Delta t$) dependent transition probability matrix can be converted to the more general matrix of rate constants **R**:

$$\mathbf{R} = \frac{1}{\Delta t} \ln(\mathbf{K}). \tag{1}$$

The corresponding emission functions $p(\mathbf{x}_t|q_i)$ project each hidden state to an observable signal (schematically shown in Figure 1 b). These emission functions are probability distributions of the multidimensional signal $\{\mathbf{x}_t\}$ while the system is in the corresponding hidden state $q_i$. The likelihood $L$ is the probability, that a hidden state trajectory $\{q_t\}$ corresponds to an observed signal (25):

$$L(\{q_t\}) = \sum_{}^{n} \pi(\mathbf{x}_1 | q_1) \prod_{t=2}^{M} \mathbf{K} \, \mathrm{diag}(p(\mathbf{x}_t | q_t)), \tag{2}$$

where $M$ is the observed trajectory length, $\pi$ is a vector of *a priori* probabilities of hidden states $q_i$ for the first observation $\mathbf{x}_1$, and ***diag*** denotes a diagonal matrix derived from a vector. This formulation of the likelihood requires initial values $\pi$ (e.g. all equal), which allows including *a priori* knowledge by freely setting these priors in the calculation of the likelihood function.

The number of the likelihood $L(\{q_t\})$ calculations for the observed trajectory for a given set of parameters $\{\mathbf{K}, p(\mathbf{x}_t|q_i), \pi\}$ is equal to the number of all possible state trajectories, which is on the order of $n^M$. With datasets of $M \sim 100'000$ frames each, like in our case, this is computationally infeasible. However, for the calculation of the likelihood defined by Eq. 2 only $M$ matrix multiplications are required. This approach is commonly known as the forward-backward procedure (10). Given the parameter set and the observed signal, the most likely state sequence (Viterbi path) can be determined (10). The power of a Hidden Markov analysis relies on the fact that iterative algorithms exist maximizing the likelihood, although those approaches are only able to locally maximize the likelihood function. In this paper we used the Baum-Welch algorithm (10) as an expectation maximization (EM) method (26).

For a Markovian system state transitions are described by the time-independent value $k_{ij}$ for each transition from state $q_i$ to $q_j$. This inherently leads to state lifetime histograms with a mono-exponential decay (27) (or with a geometric distribution in the case of discrete times). The limitation to exponentially decaying state lifetime histograms can be overcome substituting one hidden Markov state with a series of several adjacent states, which all share the same emission function (28). Additional free parameters in terms of transition constants can be introduced to allow for a wide range of different designs of state lifetime histograms. A serial adjustment of Markov states leads to a shift of the maximum of the histograms to longer times, whereas a parallel placement is able to introduce additional peaks.

The form of the emission functions is usually determined by the design of the experiment. We used a two dimensional Gauss function with the mean $\mathbf{x}_i^o$ and the covariance matrix $C_i$ to describe the probability distribution of the bead positions ($\mathbf{x}$) for the states $q_i$:

$$P(\mathbf{x} | q_i) = \frac{1}{2\pi \sqrt{|C_i|}} \exp\left\{ -\tfrac{1}{2} (\mathbf{x} - \mathbf{x}_i^o)^\mathrm{T} C_i^{-1} (\mathbf{x} - \mathbf{x}_i^o) \right\}. \tag{3}$$

**F$_1$-ATPase single-molecule assay**

An F$_1$-ATPase $\alpha_3\beta_3\gamma$ sub-complex from *Bacillus* PS3 with 10 histidine residues fused to the N-terminus of subunit $\beta$ and the mutation S106C in subunit $\gamma$ was used in the experiment (22). The histidines allow the enzyme to adsorb unspecifically on an unprocessed cover glass surface. The single cysteine available in the engineered $\gamma$ subunit was biotinylated with biotin-PEAC$_5$ maleimide. The enzyme preparation was done as described (16,29). Briefly, the F$_1$-

ATPase sub-complex (20 nM) in buffer I (10 mM MOPS pH 7.0, 50 mM KCl, 1 mM $MgCl_2$) was infused into a flow chamber consisting of two Matsunami cover glasses separated by a 50 µm spacer. After 2 minutes the flow chamber was washed with buffer II (buffer I + 10 mg/ml BSA) to remove the unbound $F_1$-ATPase and to cover the bare glass surface with BSA. Afterwards the chamber was filled with streptavidin-coated polystyrene beads (diameter 0.56 µm, Bangs Labs, ~200 pM) in buffer II. One bead per γ subunit of the $F_1$-ATPase sub-complex was bound through the stable biotin-streptavidin complex. To induce the rotation of the γ subunit of $F_1$-ATPase the flow chamber was infused with 1 mM ATP in buffer II. A scheme of the assay is presented in Figure 2. During the experiment varying concentrations of ATP/ADP/$P_i$ were applied according to table 1, with always the same $F_1$-ATPase molecule being observed. The bead motion was registered with a LOGLUX® i5 CL camera (Kamera Werk Dresden, Germany) installed on an Olympus IX71 microscope (Japan). Bead rotation was recorded with 5 ms time resolution. From the video sequences the center of mass of the rotating bead was estimated for further HMM analysis using Matlab 7.1 (MathWorks, Natick, USA).

**Results and Discussion**

By varying the concentrations of the nucleotides in the buffer solution, the $F_1$-ATPase motor was driven at different speeds. Figure 3 a shows the accumulated angle trace of the rotating bead on a single $F_1$ ATPase molecule as a function of time at seven different concentrations of ATP, ADP, and $P_i$ (see Table 1). All the measurements were done on the same $F_1$-ATPase molecule. The decrease of ATP concentration reduced the bead rotation speed in agreement with previous results (30). The plateau sections on the accumulated angle graph (Figure 3 a) were assigned to events of ADP-$Mg^{2+}$ inhibition (31,32) of $F_1$-ATPase activity and therefore stopped γ subunit rotation until ADP dissociated. The number of ADP-$Mg^{2+}$ inhibition events was not negligible at millimolar ADP concentrations (Figure 3 a).

$F_1$-ATPase has $C_3$ symmetry and the γ subunit rotates in discrete 120° steps. Figure 3 b shows a 2D density plot of the positions of the bead. The three most populated regions corresponded to the ATP hydrolysis and $P_i$ release events of the enzyme (33). These events were on average shorter than our time resolution, which resulted in an elongated dwell of the three observed stopping positions. Because the bead was connected to the $F_1$ ATPase through a $PAEC_5$ biotin-streptavidin linker, the linker flexibility introduced an additional uncertainty of ≤ 20º for the orientation of the γ subunit (31).

In order to identify the three stopping positions of the molecular motor, the simplest approach was to divide the domain of angles into three areas in a sector model. The angle boundaries were chosen in a way that the sum of squares of the distances to their corresponding mean values was minimized (black lines in Figure 3 b). Furthermore, a Markov Model with three states was applied to the data. Regarding the mean angle positions of the states and the mean values for the state durations, both approaches led to comparable results. However, the probability distribution on the boundaries of co-domains did not drop to zero (Figure 3 b), i.e. the probability distributions of the individual states overlapped and the sector model was vulnerable to mis-assignment of states in those intermediate areas.

In Figure 4, the histograms of the state durations derived from the sector model (light gray) and from the Markov model (black) are compared. The peak values for both approaches were almost at the same position (at about 60 ms), but the sector model additionally showed a high amount of very short dwells. The separation of the dwells into fluctuating events, that is, the bead moved to the second state and moved back after a short dwell, and to transition events (the bead moved to a third state after a jump to the second state) indicated that these very short dwells of the sector model belonged to the fluctuating fraction (see inset of Figure 4).

This clearly proved that those events were mis-assignments around the zone borders likely due to Brownian motion.

The results derived from the Hidden Markov Model showed no such mis-assignments. An expectation maximization algorithm searched iteratively for HMM parameters maximizing the likelihood (10). During this process an increase of the transition rates due to the fluctuating events was balanced by a decrease of the rates caused by long dwell times. Taking the fluctuating events into account. overlapping emission functions favor smaller transition rates. Whenever the bead was located near the zone borders, the contribution of the two neighboring emission functions were both comparably low. In this case, the relatively high probability to stay in the same state according to the transition matrix prevented a state change. But as soon as the bead performed a long jump, the contribution of the next emission function to the likelihood would be strongly increased and would therefore force a state change in the state trajectory (see Fig. 1 b).

After the optimal HMM parameter set was determined (state transition matrix, peak position and width of two dimensional Gaussian PDFs) the state assignment was performed and state duration histograms were built. Surprisingly, these dwell time histograms were not mono-exponential, although a single Markov state has inherently a mono-exponential state duration distribution. The observed distributions had their maxima at a position away from zero (see Fig.4, black curve). The size of the bead (~ 0.5 µm) and the flexibility of the linker introduced a latency to the system causing a further shift of the maximum to longer times in these histograms. Additionally, the intrinsic Brownian motion blurred the maxima to the observed shape.

One way to incorporate this feature of the histograms into the Markov Model is to split each Markov state into several "microstates" sharing the same emission function (28). We call a set of these microstates "macrostate" in the following. The introduced additional free parameters in terms of transition probabilities were used for a precise description of the steep left rise and the right shoulder of the state duration histograms. In the experiment, only the macrostates were visible, i.e it was only possible to assign data to macrostates. Figure 5a shows the macrostate dwell time histograms extracted from experimental data using the HMM in comparison with the state lifetime distribution found from the solution of the differential equations generated from the transition matrix. As more states were put together in a sequence, the resulting peak beaome sharper and shifted to longer times. The shape of the macrostate dwell time histograms did not change much for different models, which implicated that the state assignment was independent of the model (Figure 5 a, gray histograms).

A way to compare different model designs is to use the Bayesian information criterion (BIC) (34). The idea behind is to subtract an empirical penalty term, which should reflect the model's complexity, from the logarithm of the likelihood. Unfortunately the BIC was not sensitive to the shape of the dwell time histogram, and, therefore, an alternative model selection scheme was required. The deviation of the curves calculated using the transition matrix from the dwell time histograms based on the HMM macrostates assignment was expressed as the residual sum of squares (RSS) between both. The four types of histograms for the different macrostate transitions (that is, forward-forward, forward-backward, backward-forward, and backward-backward) were used in the RSS calculations in Fig. 5. This deviation criterion characterized much better how well the extended Markov models described the data and, in contrast to the BIC, did not require any parameters.

However, a simple alignment of multiple states in a row has one drawback: a change of the rotational direction mainly influences only the first state of a row. This is due to the higher probability of the system to remain in one microstate instead of changing it (according to the transition matrix). Hence, being in those turnaround macrostates the system will mostly remain in the first microstate of a row until the macrostate changes. This will again lead to an almost monoexponential dwell time distribution, regardless of how many microstates are in a row,(Figure 5 b).

To refine the Hidden Markov Model to overcome this drawback, additional states in the back step path had to be introduced. A similar model extension can be realized by adding a second row of states for the existing backward transitions (see Figure 5 b). In this model, turnaround events passed through two microstates and the dwell time histogram for the macrostate looked similar to the histogram for two sequential microstates shown in Figure 5a. The simplest "two row" model included just one microstate per macrostate for each direction and had four free parameters (that is, transition probabilities). Thereby only two new parameters were introduced to the model although the number of microstates had doubled (Fig. 6 a). Figure 6 b shows the superiority of the "two row" over the linear model for a sufficient number of microstates. The RSS-value of the "one row" model even increased with increasing model complexity. This is caused by the intrinsic increasing minimum state duration (equal to the number of states in a row) with increasing number of states in a row. Models with too many states in a row were not capable to describe the short-living states. Figure 6 c shows the result of the "two row" model applied to seven concentrations with increasing model complexity.

This kind of model extension leds to a universal model, which was able to assign the macrostates properly and simultaneously to find the transition rates for a wide range of the enzymatic conditions. The higher the model complexity, the better the shape of the dwell time histograms was reproduced (see RSS curves on Fig, 6 c) for all six nucleotide concentrations. Each of the four different routes through a macrostate had its own dwell time histogram to be fitted. The proper fit of the steep left flank required additional parameters. With four microstates per row (i.e. ten free parameters) the RSS reached its minimum, and additional microstates did not improve the matching of the dwell times. Basically, the shape of the histogram and the available statistics determined the required complexity of the model.

For example, the Hidden Markov Model with four microstates per row included the nucleotide concentration dependence in terms of its transition probabilities (Fig. 6 a). A single macrostate of the "two row" model is shown on the scheme with the upper row for the forward and lower row for the backward direction. The ratio of transition rate 1 to transition rate 2 is the ratio of the number of forward to backward steps, and it was found to be linearly dependent on [ATP] with a proportionality coefficient of $(3.93 \pm 1.66) \times 10^4$ M$^{-1}$. The rate 3, responsible for the transition through the state in the backward direction, was dependent on [P$_i$] in good agreement with previous results (33)), or on ADP concentration with a proportionality coefficient of $(3.30 \pm 1.24) \times 10^4$ M$^{-1}$s$^{-1}$. The corresponding coefficients calculated from the sector model $(2.76 \pm 0.96) \times 10^3$ M$^{-1}$ and $(6.23 \pm 2.57) \times 10^2$ M$^{-1}$s$^{-1}$ were completely different from those obtained using HMM. For the sector model e.g. the ratio of rate 1 to rate was influenced by a mis-assignment of the states near to the co-domains boundaries. The rate backward transition was model-dependent (see figures in the supplementary data).

Finally, using the macrostate assignment by HMM analysis it was possible to calculate the work done by the F$_1$-ATPase for each macrostate transition ($W_{tr}$) according to eq. 4

$$W_{tr} = \frac{k_B T}{D_\tau \Delta t} \sum_{i=1}^{m} (\Delta s_i)^2 , \qquad (4)$$

with $k_B$, the Boltzmann constant, $T$, the absolute temperature, $\Delta t$, the time resolution, $m$, the number of smoothed not-decreasing tangential displacements $\Delta s_i$ for a macrostate transition ($<m> \approx 13$, $<\Delta s_i> \approx 14.5$ nm, for detailed description see supplemental data), $D_\tau$, the diffusion constant for a tangential displacement ($D_\tau \approx 6.16 \times 10^{-14}$ m$^2$s$^{-1}$) calculated from $<\Delta s_0^2> = 2D_\tau \Delta t$, (with $\Delta s$, the mean square tangential displacement during $\Delta t$ without ATP consumption) using $\Delta s_0 \approx 24.8$ nm and $\Delta t = 5$ ms. Equation 4 was derived from the Einstein relation. The kinetic energy of the bead was ~10$^{-4}$ pN·nm for a 0.56 μm polystyrene bead, negligible in comparison to the thermal energy ($k_BT \approx 4.05$ pN·nm). The work produced by the $F_1$-ATPase seemed to be ATP concentration independent in good agreement with previous work (35). The work distributions turned out to be asymmetric and broad, indicating the importance of fluctuations in the present measurements. The mean work for the normal forward transition (*ff*) is $W_{ff} = (13.45 \pm 6.07) \cdot k_BT$. When a backward transition followed the forward one (*fb*), the work decreased to $W_{fb} = (9.20 \pm 4.32) \cdot k_BT$ which probably is due to the fact that the backward jumps mostly took place when the forward transition was not completed. For two consecutive backwards steps (*bb*) a work $W_{bb} = (8.23 \pm 4.03) \cdot k_BT$ was produced as well. The work of the backward transition was smaller then the chemical energy of ATP (18.4 $k_BT$ for concentration ratio C7, or 39.2 $k_BT$ for C1, respectively) and significantly smaller then the work of the forward transition. However the probability of the backward jumps was higher as expected for thermal fluctuations, i.e. also the backward transitions were ATP-driven.

**Summary**

A non-Markovian behavior of the conformational dynamics of enzymes can be described by a semi-Markov process. In this case a hidden Markov approach with an extended model design leads to a precise description of the conformational state transition data. Here the HMM design was extended by introducing state rows sharing the same emission function and replacing the single states of the standard HMM. The similarity between the dwell time histograms based on a macrostate assignment and the histograms calculated from the transition probabilities was used as a new criterion to evaluate the model quality.

We have presented a method how this approach can be applied to single-molecule experiments. To obtain a representative data set of conformational dynamics, a single immobilized $F_1$ ATPase sub-complex was driven at different speeds by altering the catalytic conditions. The extension of HMM including a row of states for each rotational direction reproduced the state duration histograms very well, while a single row model was unable to describe the histograms properly. The extended HMM approach revealed (1), an [ATP] dependence of the ratio of forward to backward jumps, (2), the [$P_i$] dependence of the rate in backward direction, (3) a likely [ATP] dependence for a backward step, and (4) a reduced work for the transition preceeding a backward step. In general this approach can be applied to a whole wealth of single-molecule systems, where a mathematical procedure is needed to identify individual states and state transitions of the molecule. This may be of importance in a field where the signal-to-noise ratio usually can only be moderately improved by averaging. Finally we would like to note that this approach may be applied to the estimation of stochastic entropy production (36) by this molecular motor.

**Acknowledgments**

The authors want to thank Kevin Murphy who provided a powerful software package for Bayesian learning algorithms as free Matlab code. This work was supported in part by the Landesstiftung Baden-Württemberg in the 'Network of Competence: Functional Nanodevices'.

**Notes**

The measurement data used in this paper as well as the Matlab source code for the HMM analysis are available upon request.

|  | [ATP], mM | [ADP], mM | [P$_i$], mM |
| --- | --- | --- | --- |
| C 1 | 1 | 0.001 | 0.001 |
| C 2 | 0.99 | 0.01 | 0.01 |
| C 3 | 0.9 | 0.1 | 0.1 |
| C 4 | 0.5 | 0.5 | 0.5 |
| C 5 | 0.1 | 0.9 | 0.9 |
| C 6 | 0.01 | 0.99 | 0.99 |
| C 7 | 0.001 | 1 | 1 |

**TABLE 1** Concentrations of the nucleotides ATP, ADP and P$_i$ used in the F$_1$-ATPase rotation experiment. The ATP to ADP and P$_i$ ratio covered six orders of magnitude.

**Figure legends**

**FIGURE 1. Scheme of a three-states Markov model. (a)** The three gray circles are the Markov states $S_1$, $S_2$ and $S_3$ with transition probabilities $k_{ij}$ marked as black arrows. **(b)** The emission functions $p(\mathbf{x}|k)$ are probability density functions of the observed signal **x**. Markov states with overlapping emission functions are not straight forward distinguishable and are therefore called hidden states.

**FIGURE 2. Experimental system**. An $F_1$-ATPase sub-complex was immobilized on a coverslip and a biotinylated bead was attached to the γ-subunit of the $F_1$-ATPase using biotin-streptavidin bonds. ATP-driven bead motion was registered using a fast camera installed on a bright field microscope.

**FIGURE 3. ATP-driven bead rotation on $F_1$-ATPase. (a)** Accumulated angle of rotation over time for the same $F_1$-ATPase molecule at different nucleotide and phosphate concentrations according to Table 1. **(b)** two dimensional density plot of the bead positions during γ subunit rotation in the presence of 1 mM ATP, 1 μM ADP and 1 μM $P_i$. The black lines mark the boundaries of the sector model. The PDFs of the three stopping positions overlap leading to a mis-assignment of the states by the sector model (see text).

**FIGURE 4. Dwell time histograms of one state** according to the sector assignment approach (light gray) and the Hidden Markov Model (black) at 1 mM ATP, 1 μM ADP and 1 μM $P_i$. **Inset**, separation of fluctuating events from transition events discloses the mis-assigned states of the sector model as fluctuating events near the borders of a state (see text for details).

**FIGURE 5. (a) Dwell time histograms of the visible state (macrostate)** assigned by HMM analysis (gray area) and calculated from the transition matrix of the same HMM (black curve) at 1 mM ATP, 1 μM ADP and 1 μM $P_i$. Each macrostate (gray ellipse) consists of 1 to 4 microstates (black circles) in a row.
**(b)** Dwell time histograms of the macrostates after the forward and before the backward transition assigned by HMM analysis (gray area) and calculated from the transition matrix of the same HMM (black curve) at 0.01 mM ATP, 0.99 mM ADP and 0.99 mM $P_i$.

**FIGURE 6 "Two row" model and residual sum of squares (RSS)** between macrostate dwell time distributions obtained for different HMMs and calculated using the HMMs transition matrices. **(a)** Scheme for a single macrostate in a "two row" model with the upper row for the forward and lower row for the backward direction. **(b)** Comparison of "one row" (●) and "two row" (∇) HMM designs for a nucleotide concentration 0.5 mM ATP, 0.5 mM ADP and 0.5 mM $P_i$. **(c)** Dependence of the RSS from the number of free parameters for different concentration ratios of the nucleotides and phosphate (C1 – C7, see table 1). The divergence between these seven curves vanishes for a sufficient complexity of the model.

**Figure 1**

a 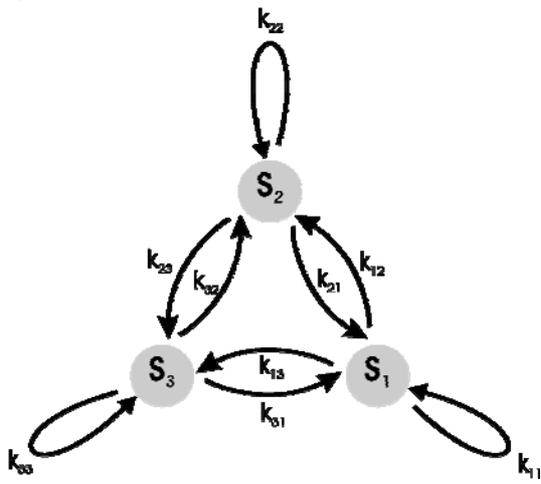 b 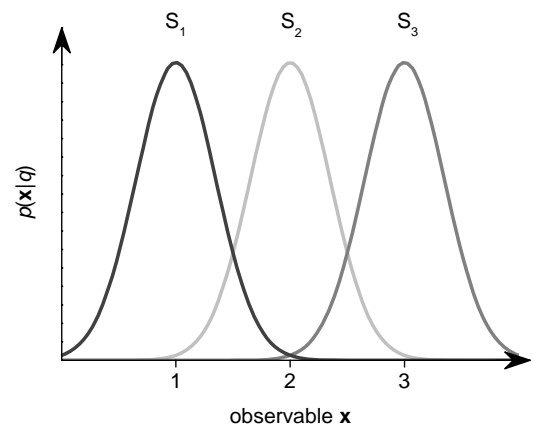

**Figure 2**

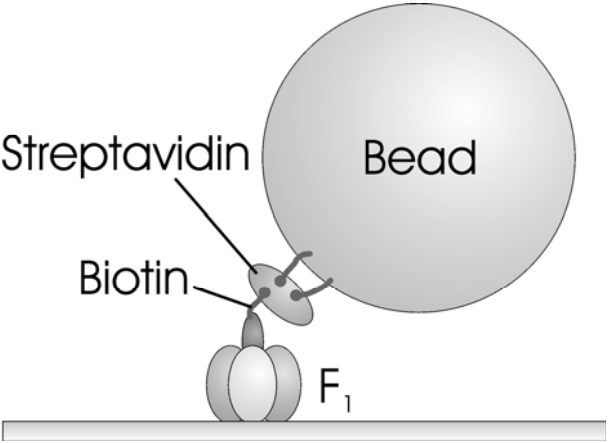

**Figure 3**

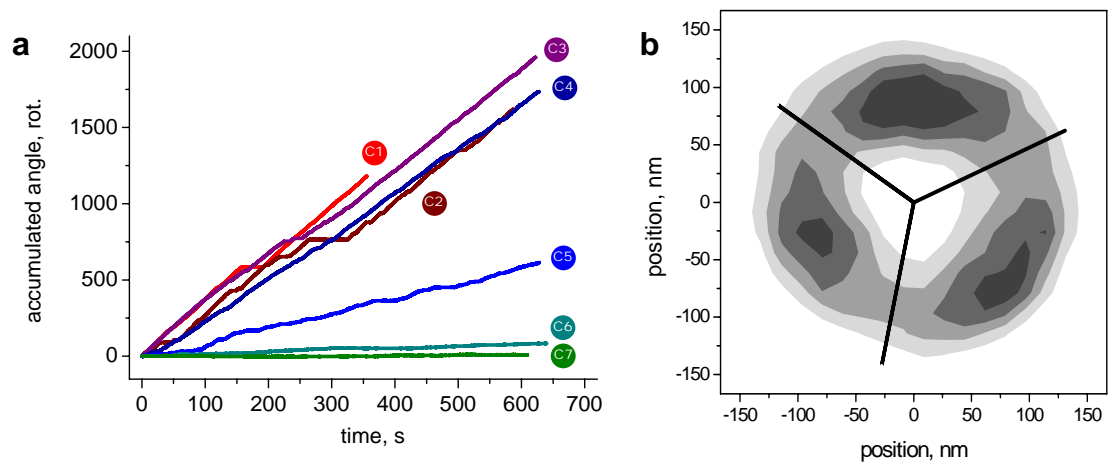

**Figure 4**

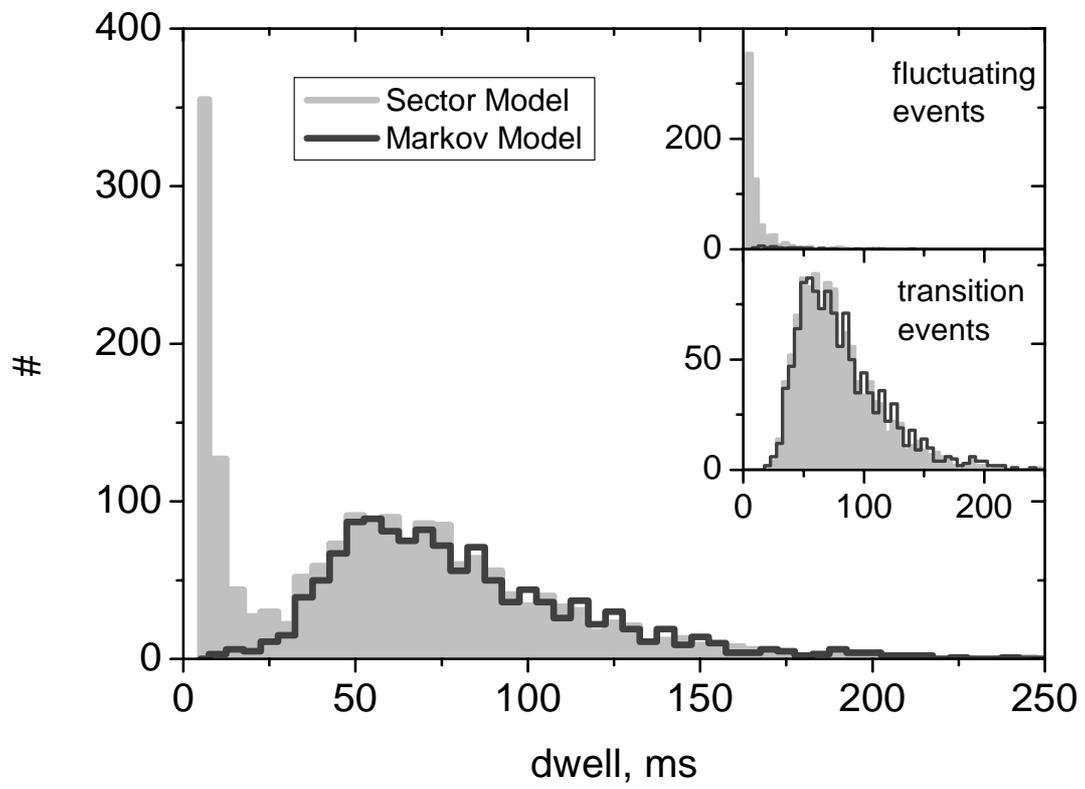

**Figure 5**

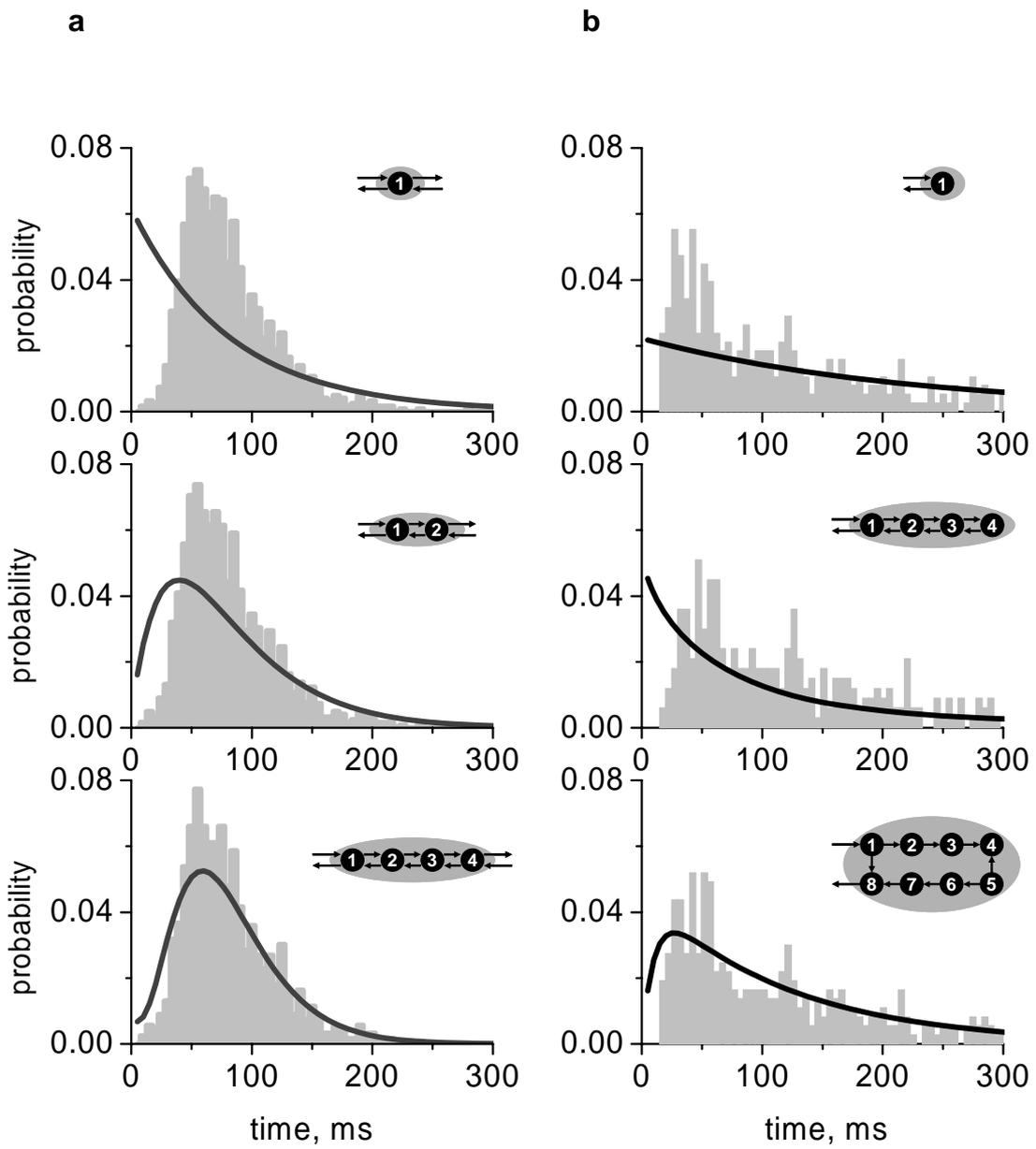

**Figure 6**

**a**

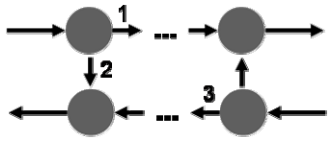

**b**
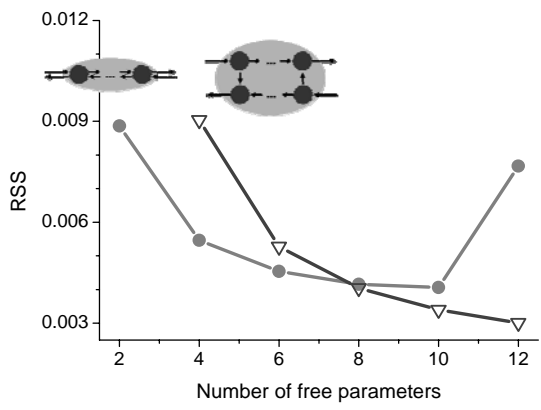

**c**
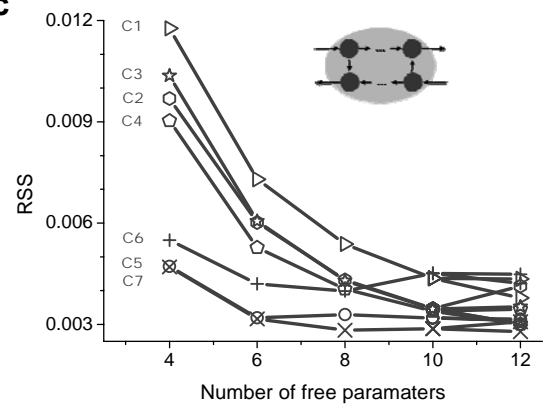

## Supplementary material

to article: **Hidden semi-Markov Models for Single-Molecule Conformational Dynamics**
A. Kovalev[*], N. Zarrabi[*], F. Werz[*], M. Börsch[*], Z. Ristic[#], H. Lill[#], D. Bald[#], C. Tietz[*], J. Wrachtrup[*]

## 1. Concentration dependence of state transition rates

HMM analysis of the bead motion delivered a state trajectory and a transition probability matrix. The transition rates between eight states ("two row" model) were calculated from the state transition probability matrix according to Eq. 1. The rates were further averaged over three $F_1$-ATPase macrostates. Dependence on nucleotide concentrations for some rates is presented in supplementary Figure S1. Since identical ADP and $P_i$ concentrations were used in the experiments (Tab. 1), we could not distinguish between ADP or $P_i$ dependence of rate 3, (see Figure S1 b).

## 2. Computation of the work produced by the $F_1$ ATPase subcomplex at ATP hydrolysis

Estimation of the mechanical work produced by the $F_1$-ATPase along with its enzyme kinetics is crucial for a thermodynamic characterization of the ATP-driven single molecule motor. Often it is assumed that the work produced by $F_1$-ATPase during ATP hydrolysis is proportional to the rotational angle of the attached bead. However, for the large beads used in our experiments, this assumption is not fulfilled. Only the tangential displacement of the bead reflects the work produced by the ATPase. To demonstrate this we first introduce this tangential displacement, then we show a way to exclude the influence of the Brownian motion on the calculation of the work, and finally the work produced by the $F_1$-ATPase is derived from these data.

Tangential displacements were found according to the scheme in Figure S2a. First, the "rotation center" (O) was defined as a center of the circle best-suited to the bead centroid positions. Then, for the two positions of the bead centroid labeled 1 and 2, the tangential displacement 1'-2' was found as a projection of the vector 1-2 on the axis perpendicular to the line going through the "rotation center" O and the middle point of the vector 1-2, (see Figure S2a). If the bead motion is just a rotational motion, the tangential displacement should be proportional to the distance of the bead from the center of rotation. However, Figure S2b demonstrates a very weak dependence of the average tangential displacement on the distance from the rotation center. Therefore the bead motion is mainly a translational motion. The friction caused by the bead rotation is much smaller than that caused by the translational motion of the bead.

The elementary work ($W_{\Delta t}$) produced by $F_1$-ATPase within a time resolution interval ($\Delta t=0.05$ s) was calculated according to

$$W_{\Delta t} = \gamma \frac{|\Delta \tilde{s}|}{\Delta t} \Delta \tilde{s}, \tag{1s}$$

where $\gamma$ is a translational hydrodynamic friction coefficient, $\Delta t$ is the time resolution, $\Delta \tilde{s}$ is a smoothed not-decreasing tangential displacement (see below). The friction coefficient was defined using the Einstein relation

$$D_\tau = \frac{k_B T}{\gamma}, \tag{2s}$$

where $k_B$ is a Boltzmann constant, $T$ is an absolute temperature, $D_\tau$ is a diffusion coefficient for the tangential motion. Combining Eq. 1s and Eq. 2s and summing up the elementary works along a sub-trajectory of the transition through a macrostate we get Eq. 4. The diffusion coefficient $D_\tau$ was determined from the longest sub-trajectory with no macrostate change (at the lowest ATP concentration) using the Einstein equation for one-dimensional diffusion and $\Delta t=0.05$ s

$$\langle \Delta s^2 \rangle = 2D_\tau \cdot \Delta t, \tag{3s}$$

where the brackest $\langle \ \rangle$ denote averaging.

Brownian motion of the bead complicated the precise estimation of the work in two ways. it affects the precision of Equation 3s, and it increases the accumulated tangential displacement (ATD) over time even without ATP consumption.

Equation 3s is valid for the classical Brownian motion of the bead. At the same time, the bead exhibits restricted diffusion, as revealed by the existence of a plateau in the curve of the mean square tangential displacement of the bead (Figure S3a). Thus we need to model the motion of the bead to find out its true diffusion coefficient. Simulations of the bead motion were done according to the stochastic differential equation

$$\gamma \Delta \mathbf{x} = \frac{\partial U_{x,y}}{\partial \mathbf{x}} \Delta t + \sqrt{4\mathbf{D}\Delta t}\ \mathbf{\eta}, \tag{4s}$$

where $\mathbf{\eta}$ is a normal distributed random vector with zero mean and variance equal to one, $\mathbf{D}$ is a diffusion matrix, $\mathbf{x}$ is a bead position vector (x,y), the bead is moving in a two dimensional potential ($U_{x,y}$). The potential in equilibrium (that is, no change of the macrostate) is determined by $U_{x,y} = -k_B T \ln(p_{x,y})$, where $p_{x,y}$ is the probability for the bead to be at position (x,y). The potential is shown on Figure S4a. Simulations showed that the relative error of the diffusion coefficients is less then 2 percent defined using Eq. 3s for the smallest $\Delta t=0.05$ s

The second challenge is showsn in Figure S3b. The average difference between the maximum and minimum accumulated tangential displacement (ATD) reaches 0.2 μm for the same macrostate (no ATP consumption) over a time interval of ~0.26 s. This is similar to the average ATD for a transition through a macrostate in the forward direction (that is, upon ATP-driven motion). Therefore at observation times >0.26 s, apparent state transition events due to Brownian motion could be incorrectly assigned to the events of the ATP-dependent rotation.

For the calculation of the work produced by the enzyme using Equation 1s the following aspects have to be considered: a transition from one macrostate to another normally takes place in many (i.e. more then five) sub-steps; the sub-step size is comparable with the step size of the Brownian motion; the energy released from ATP hydrolysis could be used only for an one-directional motion. Figure S4b (squares) shows a typical course of the ATD. To exclude thermal fluctuations the ATD curve was smoothed (Fig. S4b, dotted line). The active motion of the enzyme was assumed to be one-directional, hence the smoothed curve was further transformed to a non-decreasing curve (Figure S4b, gray line). In this way fluctuations of the enzyme potential were excluded from the work calculations. The non-decreasing smoothed ATD for each time interval were used for the work calculation according to Eq. 4.

The work for backward jumps following forward ones cannot be calculated in this way, because the bead continues its forward motion at first and only then changed the motion direction. The average ATD in this case is 0.045 μm, whereas the average difference between maximal and minimal ATD is 0.139 μm.

Typical work histograms are shown in Figure S5 for different types of transitions through a macrostate. They are broad, their shape is a gamma function in first approximation. The work was calculated for a sequence of jumps. The first step in a sequence is an initializing step showing the current microstate. The actual work was calculated for the second step. The reason of introduction of the third step is explained in the main text.

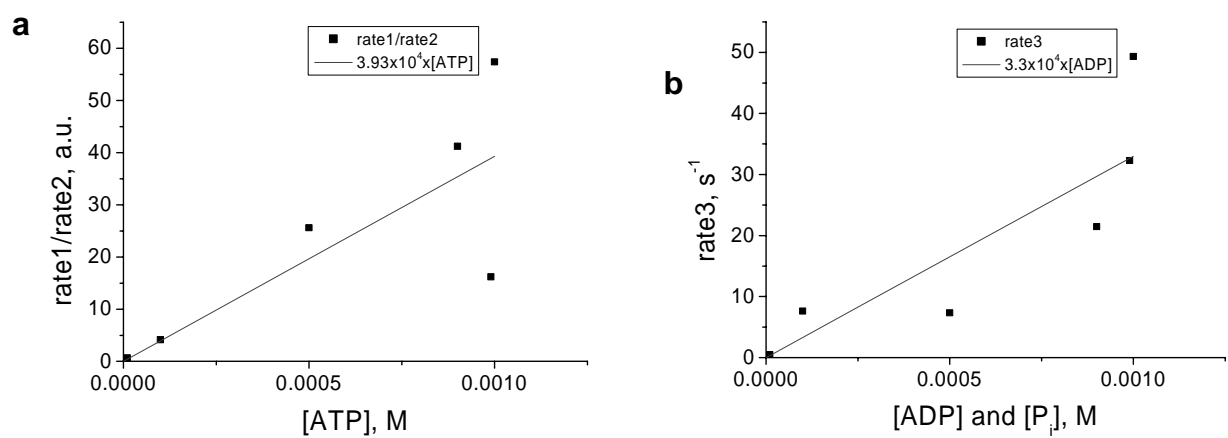

**FIGURE S1** Concentration dependence of the ratio of rates 1 and 2 (**a**), and of rate 3 (**b**). The proportionality coefficients for the linear fits are $3.93\times10^4$ [$M^{-1}$] in (a), and $3.3\times10^4$ [$s^{-1}M^{-1}$] in (b). The scheme shown in Fig. 6a in the text explains the rates numbering. Straight lines are the linear fits to the data.

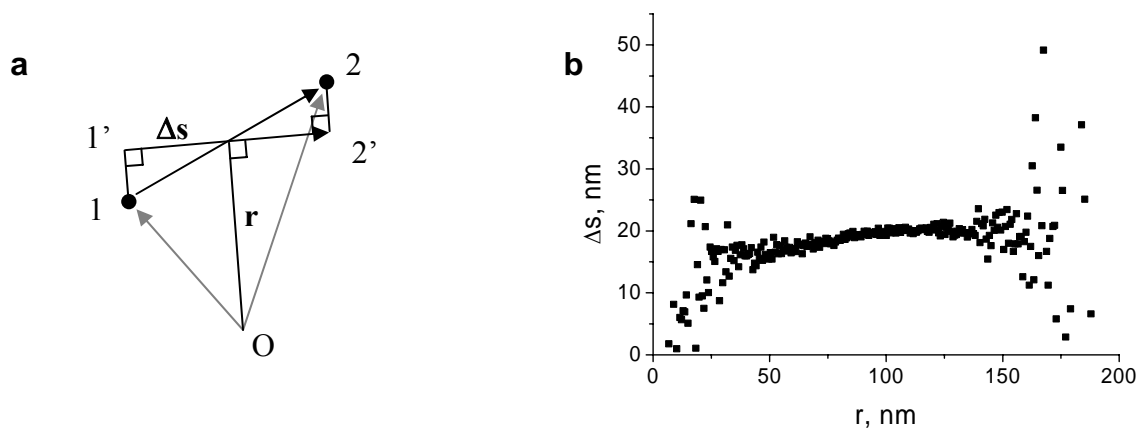

**FIGURE S2** (**a**) Scheme of the centroid of the bead motion from position 1 to position 2. 'O' is a rotation center, $\Delta\mathbf{s}$ is a tangential displacement, 1-1' and 2'-2 are radial displacements. (**b**) The tangential displacement ($\Delta s$) of the bead within 0.05 s at different radial distances (r) in the presence of 0.5 mM ATP, 0.5 mM ADP and 0.5 mM $P_i$.

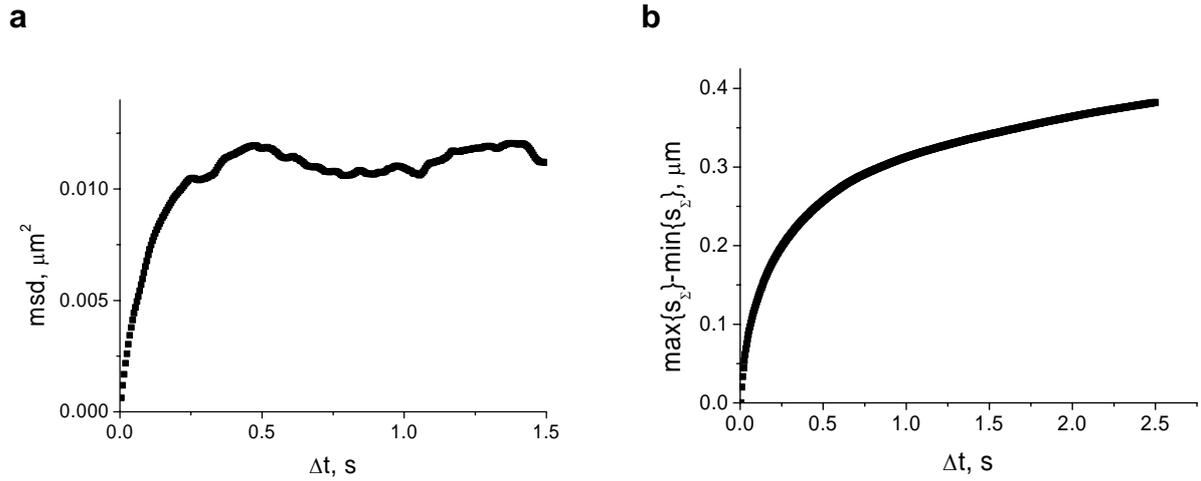

**FIGURE S3** (**a**) Dependence of the mean square of the tangential displacement of the bead on the observation time interval (Δt). (**b**) Dependence of the average maximal dispersion of the tangential displacements on time interval Δt, for nucleotides concentrations 1 μM ATP and 1 mM ADP. The dependencies were built for trajectory sections without a macrostate change.

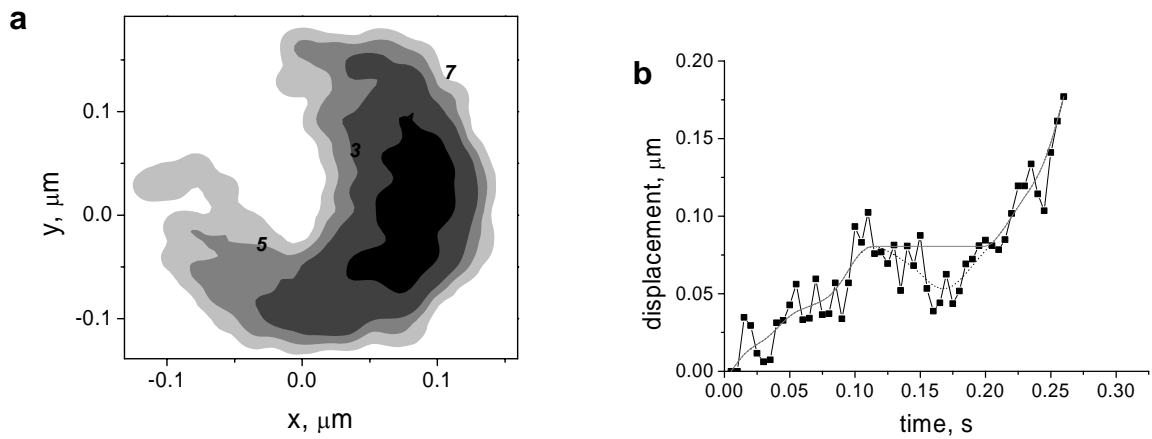

**FIGURE S4** (**a**) two dimensional potential probed by the bead on $F_1$-ATPase. The numbers on the boundaries of grayscale levels are the potential in units of $k_BT$. Concentrations of nucleotides were 1 μM ATP and 1 mM ADP. The potential was found for trajectory sections with no macrostate change. (**b**) Accumulated tangential displacement over time (squares) for a forward transition through a macrostate. The dotted line is a smoothed ATD curve, the gray line is a non-decreasing smoothed ATD curve.

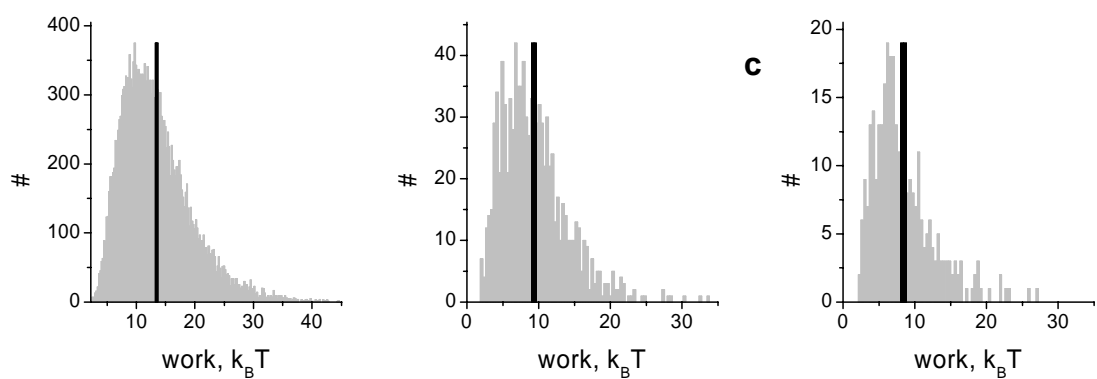

**FIGURE S5** Histograms of works produced by $F_1$-ATPase for: (**a**) a forward transition after a forward transition before the next forward transition (fff), (**b**) a forward transition after a forward transition before a backward transition (ffb), and (**c**) a backward transition after a backward transition (bb). Black vertical lines show the mean values.